# Preserving Twist-Angle in Marginally Twisted Double-Bilayer Graphene Devices During Fabrication

*Hyeon-Woo Jeong[1], Jiho Kim[2], Boknam Chae[2], Kenji Watanabe[3], Takashi Taniguchi[4] and Gil-Ho Lee[1,*]*

[1] Department of Physics, Pohang University of Science and Technology, Pohang 37673, Korea

[2] Pohang Accelerator Laboratory, Pohang University of Science and Technology, Pohang 37673, Korea

[3] Research Center for Electronic and Optical Materials, National Institute for Materials Science, 1-1 Namiki, Tsukuba 305-0044, Japan

[4] Research Center for Materials Nanoarchitectonics, National Institute for Materials Science, 1-1 Namiki, Tsukuba 305-0044, Japan

**ABSTRACT**

Twisted van der Waals heterostructures provide a platform for studying a wide range of electron correlation phenomena, including unconventional superconductivity and correlated insulating states. However, fabricating such devices is challenging due to the difficulty in achieving and maintaining homogeneous twist-angles. Here, we present a fabrication method to preserve the twist-angle with minimal deformation. We fabricated marginally twisted double-bilayer graphene (mTDBG) stacks and directly imaged the resulting triangular superlattice periodicity via scattering-type scanning near-field optical microscopy (s-SNOM). This technique enabled us to monitor twist-angle deformation at each fabrication step, paving the way for more reliable device fabrication and facilitating the exploration of twist-angle-dependent physics.

**Introduction**

Twisted graphene systems have been known to exhibit exotic phases[1-10] driven by strong electron correlations. At specific twist-angles, these systems develop flat bands where kinetic energy is quenched, amplifying the effects of electron-electron interactions, leading to phenomena such as superconductivity[1, 6, 8], Mott insulator behavior[2, 10] and anomalous magnetic[4, 7] and electric[3] properties. At small twist-angles, the Moiré lattice expands, leading to atomic reconstruction[11]. Previous imaging studies have revealed that marginally twisted graphene systems undergo atomic reconstruction[11], forming alternating stacking orders (Bernal and rhombohedral) and domain walls to minimize regions with AA-type stacking[11-15]. While the twisted graphene systems provide an excellent platform for investigating electron-electron interactions, device fabrication presents considerable challenges, primarily due to the twist-angle inhomogeneity across the device[6]. Maintaining a uniform twist-angle throughout the stacking and fabrication processes is crucial for fully realizing the potential of twisted graphene systems in electronic applications.

This study addresses these challenges by developing fabrication techniques that preserve the twist-angle while minimizing deformation in the lattice structure. Here, we report our findings on how fabrication processes impact the triangular lattice structure of marginally twisted double-bilayer graphene (mTDBG). Scattering-type scanning near-field optical microscopy (s-SNOM)[16], a powerful imaging tool with nanoscale resolution, was employed to observe the triangular lattice domains of mTDBG directly. By systematically tracking changes in the lattice structure throughout key fabrication steps—including stacking, etching, and metal deposition—we sought to identify and mitigate factors contributing to the deformation of the triangular lattice.

**Experimental Methods**

The hBN and graphene flakes were exfoliated onto Silicon substrates with a 280 nm oxide layer. All stacks used in this experiment were dry-transferred using polypropylene carbonate (PPC)/polydimethylsiloxane (PDMS) stamps. When surface impurities on the stack interfered with imaging, annealing was performed in a vacuum at a temperature range of 350–500 °C for 2 hours to remove the impurities. We identified the triangular lattice using s-SNOM (Neaspec neaSNOM) combined with an infrared QCL laser (Daylight Solutions).

The electrodes and etching masks were fabricated using standard electron-beam lithography techniques. The hBN and graphene layers were selectively etched with a reactive ion etcher (RIE) under $CF_4$ and $O_2$ plasma, respectively. Cr (5 nm) and Au (100 nm) layers were deposited onto the exposed contact areas using an electron-beam evaporator.

**Results and Discussion**

Bernal and rhombohedral stacking structures (Figure 1A, 1B) in tDBG can be distinguished by differences in their infrared optical contrast due to their distinct electronic structures[17, 18]. We identified the triangular lattice of mTDBG using infrared s-SNOM[16], a technique offering a spatial resolution of a few tens of nanometers. Initially, a mTDBG stack was prepared using the tear-and-stack method[19], incorporating a large top hBN layer that covered the entire mTDBG area. Figure 1C illustrates the flipping process to orient the mTDBG side upward, preventing solvent or polymer contamination. This process involved spin-coating a polypropylene carbonate (PPC) film onto the top of the stack, detaching the stack from the original substrate by peeling off the PPC film and transferring the inverted stack onto a new substrate. Subsequent annealing removed the PPC film, resulting in a clean mTDBG surface facing upward on the hBN crystal. Figure 1D shows an optical microscopy (OM) image of the mTDBG stack following the fabrication process depicted in Figure 1C. The topography in Figure 1E, acquired simultaneously with the s-SNOM phase image, reveals no distinct features. In contrast, the third-harmonic phase image of the mTDBG taken by s-SNOM (Figure 1F) reveals the triangular lattice and linear domain boundaries. The area exhibiting a uniform twist-angle represents only a small fraction of the total mTDBG area. Therefore, s-SNOM imaging is essential for fabricating devices with a consistent twist-angle. To enhance twist-angle uniformity for electronic device applications, we employed s-SNOM to investigate changes in the lattice structure occurring during the fabrication processes, including stacking, etching, and metal deposition.

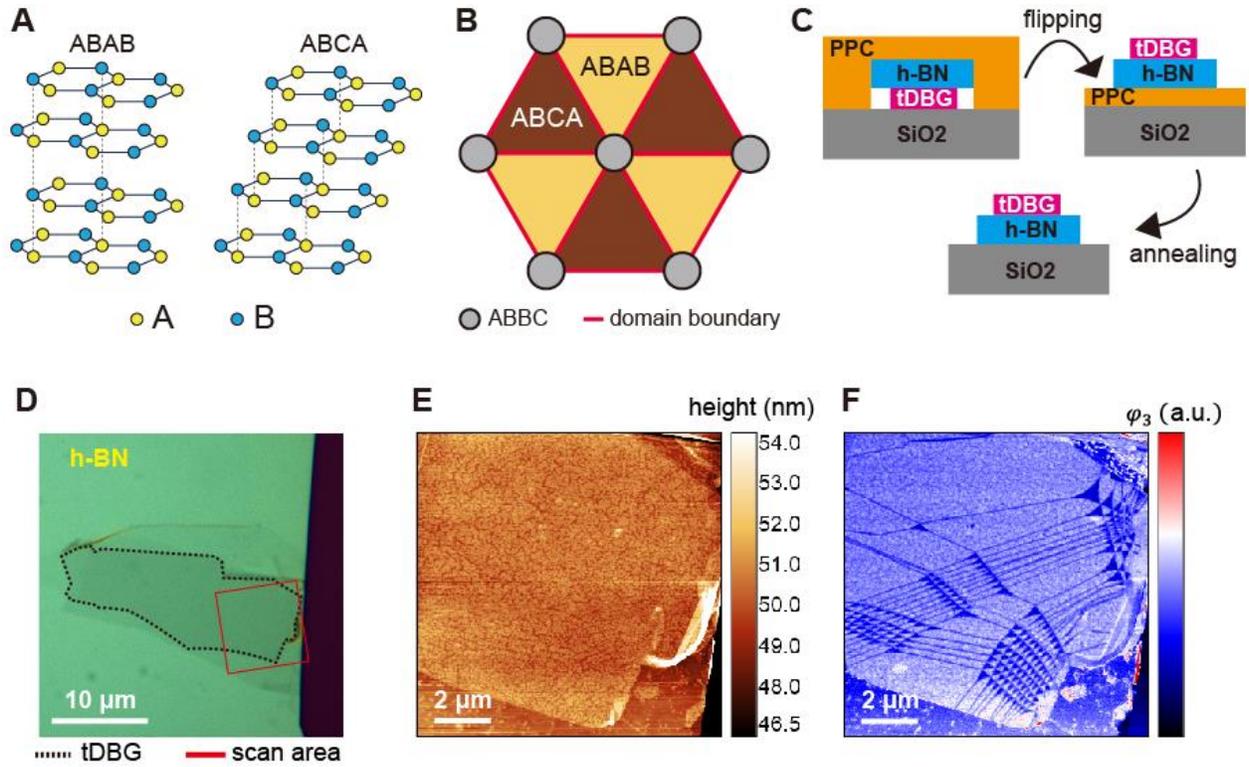

**Figure 1. Triangular lattice of marginally ($\theta_{twist}$ < 0.1 °) twisted dual-bilayer graphene (mTDBG).** (A) Bernal (ABAB) and rhombohedral (ABCA) stacking structure of tetralayer graphene. (B) Alternative stacking orders with domain boundary in mTDBG. (C) Flipping process of the stack consists of hexagonal boron nitride(hBN) and tDBG using polypropylene carbonate (PPC) film. (D) Optical image of mTDBG. (E) Topographic image of the area outlined by the red dotted box in (D) acquired using atomic force microscopy. (F) Third-harmonic phase ($\varphi_3$) image of the same area of (E) acquired using scattering scanning near-field optical microscopy.

To protect the mTDBG on the bottom hBN from surface contamination during subsequent fabrication, we covered it with a top hBN layer. A thin top hBN layer, approximately 3 nm thick, was chosen to enable imaging of the underlying mTDBG. During the covering process, the top hBN was released onto the mTDBG stack at 110°C. Given that the sample had already been annealed at 350 °C (Figure 1C), we assumed minimal deformation of the triangular lattices upon heating to 110 °C (see Supporting Information, Figure S1). We monitored the mTDBG area by acquiring s-SNOM images before and after the top hBN transfer. A solid red line delineates the area exhibiting triangular lattices in the s-SNOM image before top hBN transfer (Figure 2A). After covering the sample with the top hBN, the triangular lattices exhibited significant deformation from their initial configuration (Figure 2B). We hypothesize that the mechanical strain applied during the release of the top hBN deformed the triangular lattice of the mTDBG. To investigate this, we performed the bottom and top hBN covering processes separately to assess the deformation induced at each stage. However, because minimizing process repetitions reduces deformation, a more practical approach would be to fully encapsulate the stack initially, release it onto the substrate, verify the twist angle, and proceed with subsequent fabrication steps.

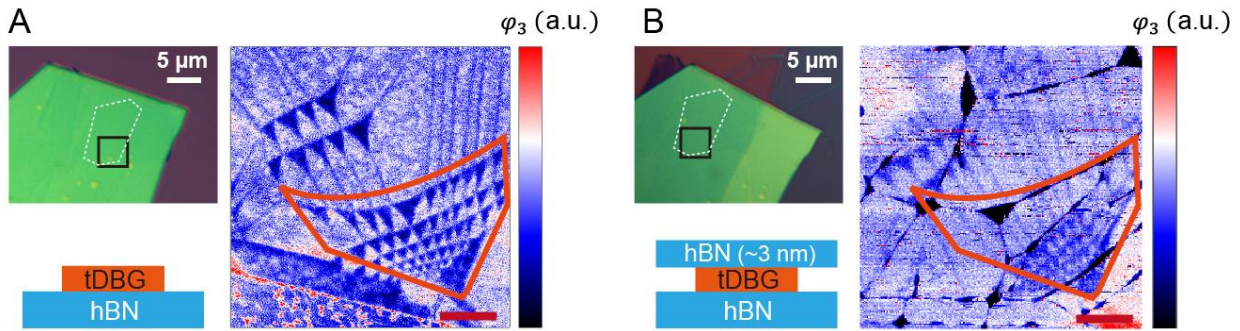

**Figure 2. Lattice deformation during dry-transfer process.** (A, B) Optical images of the mTDBG stack (upper left), third-harmonic phase images of the triangular lattice in the mTDBG

stack acquired using infrared s-SNOM (right), and device schematics (lower left) (A) before and (B) after covering with top hexagonal boron nitride. The white dashed line delineates the mTDBG area, and the black box indicates the s-SNOM imaging region. The triangular lattices are highlighted by the red line. The scale bar represents 1 $\mu$m.

Reactive ion etching of hBN and graphene is necessary for shaping devices into a specific geometry. To investigate how the etching of sample boundary induces deformation in the triangular lattices, we fabricated a fully encapsulated mTDBG stack with top and bottom hBN (Figure 3A). The initial triangular lattices of the mTDBG stack before etching are shown in the right panel of Figure 3A. Following s-SNOM imaging, we fabricated an etching mask of a standard Hall-bar geometry, outlined by the yellow line in Figure 3A, using electron-beam lithography. Subsequently, we etched the hBN and graphene outside this mask. The s-SNOM image in the right panel of Figure 3B shows that the triangular lattices were deformed after the etching process. As the stack was neither heated above room temperature nor subjected to external mechanical strain from the etching mask lithography through to the end of the etching process, the observed deformation of the triangular lattices is attributed solely to the boundary etching of the stack. The triangular lattices are known to exhibit intrinsic strain between adjacent triangles[14]. Consequently, etching a portion of mTDBG can trigger local strain relaxation, reversing Bernal stacking. As shown in the right panel of Figure 3B, the triangular domains near the etching edge disappeared and relaxed to Bernal stacking. The etching boundary must be positioned sufficiently far from the original triangular lattice regions to preserve the original triangular lattices.

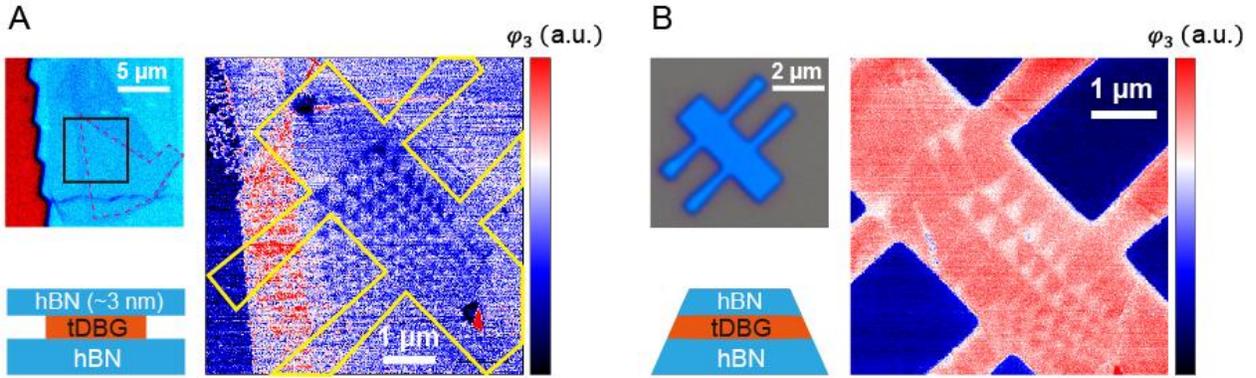

**Figure 3. Deformation of triangular lattices after the boundary etching process.** (A, B) Optical images of the mTDBG stack (upper left), third-harmonic phase images taken with infrared s-SNOM of the triangular lattice in the mTDBG stack (right), and device schematics (lower left) (A) before or (B) after reactive ion etching into a Hall-bar shape outlined by yellow line. The red dashed and black solid lines in (A) outline the mTDBG region and the s-SNOM imaging area, respectively.

Metal contact deposition can also induce strain in graphene devices[20]. Here, we demonstrate a non-invasive electrical contact method that minimizes the deformation of the triangular lattices. Conventional one-dimensional (1D) contacts to the graphene layer[21] involve etching the graphene to expose a 1D edge, followed by contact metal deposition to establish electrical contact. However, as discussed in relation to Figure 3, triangular lattice deformation can occur following graphene etching. To avoid this etching process, we selectively etched only the top hBN layer, leaving the mTDBG layer intact instead of etching the entire hBN/mTDBG/hBN heterostructure. This exposed the mTDBG layer contact area, enabling direct deposition of Cr (5 nm)/Au (80 nm) contact electrodes onto the mTDBG layer (Figure 4B) to form two-dimensional (2D) contacts. Subsequently, the sample was annealed at 350 °C to remove surface contaminants before s-SNOM imaging. In contrast to the samples with edge-contacted electrodes (Supporting Information, Figure S2), the initial triangular lattices remained virtually unchanged after surface contact, as shown in the right panel of Figure 4B (Supporting Information, Figure S3).

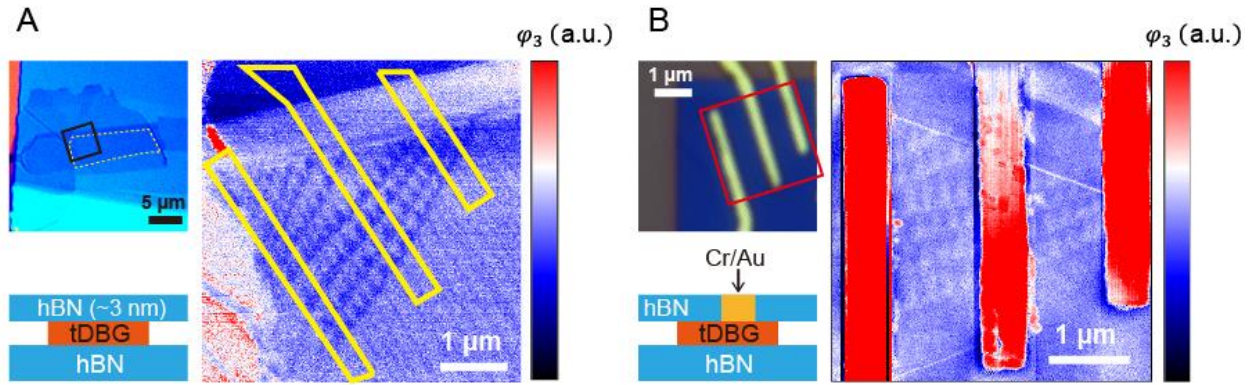

**Figure 4. Triangular lattice with minimal deformation after metal deposition.** (A, B) Optical images of the mTDBG stack (upper left), third-harmonic phase images of the triangular lattice in the mTDBG stack acquired using infrared s-SNOM (right), and device schematics (lower left) (A) before and (B) after 2D contact metal deposition. The electrode design is outlined by yellow lines in the s-SNOM image in (A). The yellow dashed line in (A) outlines the mTDBG. The black solid line in (A) and the red solid line in (B) outline the s-SNOM imaging area.

**Conclusion**

Using s-SNOM to monitor the triangular lattice formed in marginally twisted double bilayer graphene (mTDBG), we investigated lattice deformations induced by various device fabrication processes, including the dry transfer of van der Waals layers, reactive ion etching, and deposition of electrical contact metals. Our findings revealed that the intrinsic strain within the triangular lattice renders it highly susceptible to deformation during mTDBG layer etching, leading to a shift toward zero twist-angle and partial reversion to Bernal stacking. We developed a non-invasive fabrication method to mitigate this deformation that avoids mTDBG layer etching by employing 2D metal contacts. This approach effectively preserved the triangular lattice throughout the fabrication process and maintained the initial lattice configuration even after annealing at temperatures up to 350 °C to remove surface contaminants. These results offer valuable guidelines for optimizing fabrication processes for strain-sensitive graphene-based devices and improving twist-angle uniformity and device performance, which are crucial for advancing the study of twisted graphene systems.

ASSOCIATED CONTENT

**Supporting information:** Supplementary Information contains s-SNOM images showing the effect of annealing temperature, lattice deformation with 1D edge contacts, and lattice deformation with 2D surface contacts.

AUTHOR INFORMATION

**Corresponding Author**

* Correspondence to Gil-Ho Lee (lghman@postech.ac.kr)

**Author Contributions**

G.-H.L. conceived and supervised the project. H.-W.J. designed and fabricated the devices and conducted the measurements. H.-W.J. performed s-SNOM imaging with assistance from J.K. and B.C. T.T. and K.W. provided the hBN crystal. H.-W.J. and G.-H.L. analyzed the data and wrote the manuscript. All authors approved the final version of the paper.

ACKNOWLEDGMENT

This work was supported by National Research Foundation (NRF) Grants (Nos. RS-2022-NR068223, RS-2024-00393599, RS-2024-00442710, RS-2024-00444725) and ITRC program (IITP-2025-RS-2022-00164799) funded by the Ministry of Science and ICT, Samsung Science and Technology Foundation (Nos. SSTF-BA2401-03 and SSTF-BA2101-06), and Samsung Electronics Co., Ltd. (IO201207-07801-01). K.W. and T.T. acknowledge support from the JSPS KAKENHI (Grant Numbers 21H05233 and 23H02052) , the CREST (JPMJCR24A5), JST and World Premier International Research Center Initiative (WPI), MEXT, Japan.